\documentclass[twocolumn,amsmath,amssymb]{revtex4}
\usepackage[dvips]{graphicx}
\usepackage{dcolumn}
\usepackage{bm}

\begin{document}
\title[]
{A single model of traversable wormholes supported by generalized phantom energy or Chaplygin gas}
\author{Peter K.\,\,F. Kuhfittig}
\address{Department of Mathematics\\
Milwaukee School of Engineering\\
Milwaukee, Wisconsin 53202-3109 USA}
\date{\today}

\begin{abstract}
\noindent
This paper discusses a new variable equation of state parameter leading 
to exact solutions of the Einstein field equations describing traversable 
wormholes.  In addition to generalizing the notion of phantom energy, the 
equation of state generates a mathematical model that combines the 
generalized phantom energy and the generalized Chaplygin gas models. 

\vspace{12pt} 
\noindent 
PAC numbers: 04.20.Jb, 04.20.Gz
\end{abstract}

\maketitle 


\section{Introduction}\label{S:Introduction}
\noindent Traversable wormholes, whose 
existence was first conjectured by Morris and Thorne in 1988 \cite{MT88}, 
may be defined as handles or tunnels in the spacetime topology linking 
different universes or widely separated regions of our own Universe.  
A renewed interest in wormholes is due in part to the discovery that 
our Universe is undergoing an accelerated expansion \cite{aR98, sP99}, 
that is, $\overset{..}{a}>0$ in the Friedmann equation 
$\overset{..}{a}/a=(-4\pi/3)(\rho+3p)$, using units in which 
$c=G=1$.  The cause of the acceleration is a negative pressure 
\emph{dark energy}, a form of matter whose equation of state is 
$p=-K\rho$, $\rho>0,$ and $K$ a constant; $p$ is the spatially 
homogeneous pressure and $\rho$ the energy density.  A value of 
$K>1/3$ is required for an accelerated expansion.  The case 
$1/3<K<1$ is referred to as \emph{quintessence}, while $K=1$ corresponds 
to a cosmological constant \cite{rB07}.  Of particular interest is the 
case $K>1$, referred to as \emph{phantom energy}, since it leads to a 
violation of the null energy condition, an essential requirement for 
maintaining a wormhole \cite{MT88}.  Matter that violates the null 
energy condition is usually called \emph{exotic}.  Since the notion of 
dark or phantom energy ordinarily applies only to a homogeneous 
distribution of matter, while wormhole spacetimes are necessarily 
inhomogeneous, phantom energy does not automatically qualify as a 
candidate for exotic matter.  It is shown in Ref. \cite{sS05}, however,   
that an extension to spherically symmetric inhomogeneous spacetimes can 
be carried out.

An alternative model is based in Chaplygin gas, whose equation of state 
is given by $p=-K/\rho$.  Another possibility is generalized Chaplygin 
gas (GCG), whose equation of state is $p=-K/\rho^a$, $0<a\le 1$ 
\cite{fL06, KMP, BBS, pK08}.  Cosmologists became interested in this 
form of matter when it turned out to be a candidate for combining 
dark matter and dark energy: in early times Chaplygin gas behaves 
like matter and in later times like a cosmological constant.  To 
support a wormhole, we must have \cite{fL06} 
\begin{equation}\label{E:urinequality}
    K<\frac{1}{(8\pi r_0^2)^{a+1}},
\end{equation}
where $r=r_0$ is the throat of the wormhole, defined below.

Two recent papers \cite{fL07, fR07} discussed wormhole solutions that 
depend on a variable equation of state parameter: $p/\rho=-K(r),$ 
where $K(r)>1$ for all $r$.  The variable $r$ refers to the radial 
coordinate in the line element
\begin{equation}\label{E:line2}
   ds^2=-e^{2f(r)}dt^2+\frac{dr^2}{1-\frac{b(r)}{r}}dr^2
    +r^2(d\theta^2+\text{sin}^2\theta\,d\phi^2).  
\end{equation}
In this form of the line element, $f=f(r)$ is called the 
\emph{redshift function}.  The minimum radius $r=r_0$ corresponds to 
the \emph{throat} of the wormhole, where $b(r_0)=r_0$.

Ref. \cite{pKnote1} assumes that $K$ is both space and time dependent, 
i. e., $K=K(r,t)$.  Evolving equations of state are also discussed in 
Refs. \cite{MRC, SS06a, SS06b}.

An interesting solution of the Tolman-Oppenheimer-Volkoff equation 
in a Chaplygin-gas setting can be found in Ref. \cite{vG08}.  Questions 
of stability are addressed in Ref. \cite{ES07}. 

In this paper we return to the form $K=K(r)$.  The purpose is to 
show that the parameter $K(r)=Br^{2(a-1)}$ in the equation of state 
leads to a unified model for the generalized Chaplygin gas and phantom 
energy wormholes, also generalized.  More precisely, the equation of 
state, a motivation for which is given in the next section, is 
\[
 p=-Br^{2(a-1)}\rho^a, \quad a\neq 0,
\]
and where $B$ is a positive constant.  As already noted, $-1\le a<0$ 
corresponds to the generalized Chaplygin case, so that, analogously, 
$a>0$ generalizes the phantom-energy case.  (Recall that the latter 
normally assumes that $a=1$).

We are striving in all cases for exact solutions.  While such 
solutions are not usually obtainable, they have the advantage of 
being relatively easy to state and to analyze, without being 
exhaustive.  On the contrary, the exact solutions suggest the 
existence of a whole class of solutions with a variable parameter 
that could model the properties described here.  To obtain exact solutions, 
the redshift function $\gamma=\gamma(r)$ in the general line element     
\begin{equation}\label{E:line3}
 ds^2=-e^{2\gamma(r)}dt^2+e^{2\alpha(r)}dr^2+r^2(d\theta^2
     +\text{sin}^2\theta\,d\phi^2)
\end{equation}
needs to be inserted ``by hand," as in Ref. \cite{pK06}.

From the Einstein field equations in the orthonormal frame, 
$G_{\hat{\alpha}\hat{\beta}}=8\pi T_{\hat{\alpha}\hat{\beta}}$,
the components of the Einstein tensor are proportional to the 
components of the stress-energy tensor.  In particular, 
$T_{\hat{t}\hat{t}}=\rho$, $T_{\hat{r}\hat{r}}=p$, and 
$T_{\hat{\theta}\hat{\theta}}=T_{\hat{\phi}\hat{\phi}}=p_t$, where 
$\rho$ is the energy density, $p$ the radial pressure, and $p_t$ the 
lateral pressure.  The weak energy condition (WEC) is given by  
$T_{\hat{\alpha}\hat{\beta}}\mu^{\hat{\alpha}}\mu^{\hat{\beta}}
\ge 0$ for all time-like vectors and, by continuity, all null vectors.  
For the radial outgoing null vector $(1,1,0,0)$, the WEC now becomes 
$\rho+p\ge 0.$  So if this condition is violated, we have $\rho+p<0.$

\section{Variable equations of state}\label{S:variable}\noindent
The first step in this section is to list the components 
of the Einstein tensor in the orthonormal frame \cite{pK02}:
\begin{equation}\label{E:Einstein1}
   G_{\hat{t}\hat{t}}=\frac{2}{r}e^{-2\alpha(r)}\alpha'(r)
      +\frac{1}{r^2}(1-e^{-2\alpha(r)}),
\end{equation}
\begin{equation}\label{E:Einstein2}
   G_{\hat{r}\hat{r}}=\frac{2}{r}e^{-2\alpha(r)}\gamma'(r)
       -\frac{1}{r^2}(1-e^{-2\alpha(r)}),
\end{equation}
\begin{multline}\label{E:Einstein3}
   G_{\hat{\theta}\hat{\theta}}=G_{\hat{\phi}\hat{\phi}}
      =e^{-2\alpha(r)}\left(\gamma''(r)-\gamma'(r)\alpha'(r)
         \phantom{\frac{1}{r}}\right.\\
      \left.+[\gamma'(r)]^2+\frac{1}{r}\gamma'(r)-\frac{1}{r}\alpha'(r)
         \right).
\end{multline}

We observe next that the general line element (\ref{E:line3}) 
can also be put into the form  of Eq.~(\ref{E:line2}):
\begin{equation}\label{E:line4}\
   ds^2=-e^{2\gamma(r)}dt^2+\frac{dr^2}{1-\frac{b(r)}{r}}dr^2
    +r^2(d\theta^2+\text{sin}^2\theta\,d\phi^2). 
\end{equation}
So
\begin{equation}\label{E:shape1}
  e^{2\alpha(r)}=\frac{1}{1-\frac{b(r)}{r}}\quad \text{and}\quad
   b(r)=r\left(1-e^{-2\alpha(r)}\right ).
\end{equation}
Since $b(r_0)=r_0$, $\alpha$ must have a vertical asymptote at 
$r=r_0$: $\lim_{r \to r_0+}\alpha(r)=+\infty,$ while $\gamma=
\gamma(r)$ must never be zero to avoid an event horizon. In 
addition, the shape function $b=b(r)$ must satisfy the flare-out 
conditions at the throat \cite{MT88}: $b(r_0)=r_0$ and 
$b'(r_0)<1.$   Another requirement is asymptotic flatness: 
$b(r)/r\rightarrow 0$ as $r\rightarrow \infty.$  Finally, using 
Eqs.~(\ref{E:shape1}) and (\ref{E:Einstein1}), one can readily 
verify that 
\begin{equation}\label{E:rho}
   \rho=\frac{b'(r)}{8\pi r^2}.
\end{equation}

The need for a variable $K$ arises quite naturally: the first 
specialization, $\gamma(r)=\frac{1}{2}\text{ln}\,\frac{c}{r}$, 
suggested by Zaslavskii \cite{oZ05}, yields for a constant $K>1$ 
\cite{pK06, oZ05} 
\begin{equation}\label{E:Zaslavskii}
   e^{2\alpha(r)}=\frac{1}{\left(1-\frac{1}{K}\right)
     \left(1-\frac{r_0}{r}\right)},
\end{equation}
an exact solution for a phantom-energy wormhole
($p=-K\rho$).  So from Eqs.~(\ref{E:Einstein2}) and (\ref{E:rho}), 
\[
   -K(r)=\frac{p}{\rho}=-\frac{1/(8\pi r^2)}
         {b'(r)/(8\pi r^2)}=-\frac{1}{b'(r)}.
\]
Hence
\[
     p=-\frac{1}{b'(r)}\rho.
\]
In particular, from Eqs.~(\ref{E:Zaslavskii}) and 
(\ref{E:shape1}), 
\[
    b(r)=r\left[1-\left(1-\frac{1}{K}\right)
       \left(1-\frac{r_0}{r}\right)\right],
\]
which yields $b'(r)=1/K$.  Since $K>1$, it follows that 
$b'(r_0)<1,$ as required.

Using the generalized equation of state, $p=-K(r)\rho^a,$ we 
obtain
\begin{multline*}
    -K(r)=\frac{p}{\rho^a}=-\frac{1/(8\pi r^2)}
      {[b'(r)]^a/[(8\pi)^ar^{2a}]}\\
       =-\frac{(8\pi)^{a-1}r^{2a-2}}{[b'(r)]^a}.
\end{multline*}
So if $b'=1/K$ again, then 
\begin{equation}\label{E:ES1}
   K(r)=K^a(8\pi)^{a-1}r^{2a-2}.
\end{equation}
As a consequence, we will assume from now on that the equation of state has the form
\begin{equation*}
    p=-Br^{2(a-1)}\rho^a,\quad a\ne 0,
\end{equation*}
where $B$ is a positive constant.  It will be seen later that $a$ can be either positive or negative. 

\section{The first two solutions}\label{S:first}
\noindent
From Sec. \ref{S:variable} the equation of state is 
given by
\begin{equation}\label{E:equationofstate}
    p=-Br^{2(a-1)}\rho^a, \quad a\neq 0.
\end{equation}
The Einstein field equations $G_{\hat{\alpha}\hat{\beta}}=
8\pi T_{\hat{\alpha}\hat{\beta}}$ now yield
\[
   \rho=\frac{1}{8\pi}G_{\hat{t}\hat{t}}\quad \text{and}\quad
    \frac{1}{8\pi}G_{\hat{r}\hat{r}}=p=[-K(r)]
       \rho^a, 
\]  
whence
\begin{multline}\label{E:equation1}
   \frac{1}{8\pi}\left[\frac{2}{r}e^{-2\alpha(r)}\gamma'(r)
     -\frac{1}{r^2}\left(1-e^{-2\alpha(r)}\right)\right]
   =-Br^{2(a-1)}\\\times\left[\frac{1}{8\pi}\left(\frac{2}{r}
      e^{-2\alpha(r)}\alpha'(r)+\frac{1}{r^2}
          \left(1-e^{-2\alpha(r)}\right)\right)\right]^a.
\end{multline}
Zaslavskii's function $\gamma(r)=\frac{1}{2}\text{ln}\,\frac{c}{r}$ 
allows us to solve Eq.~(\ref{E:equation1}) by separation of variables: 
substituting $\gamma'(r)=-1/2r,$ we get directly
\begin{multline*}
   \frac{1}{8\pi}\frac{1}{r^2}=Br^{2(a-1)}\\
       \times\left[\frac{1}{8\pi}\left(\frac{2}{r}
      e^{-2\alpha(r)}\alpha'(r)+\frac{1}{r^2}
          \left(1-e^{-2\alpha(r)}\right)\right)\right]^a.
\end{multline*}
Raising each side to the power $\frac{1}{a}$ produces
\begin{multline*}
   \frac{1}{r^{2/a}}=B^{1/a}r^{2(1-1/a)}\left(\frac{1}{(8\pi)^{1-1/a}}\right)\\
    \times\left[\frac{2}{r}e^{-2\alpha(r)}\alpha'(r)+\frac{1}{r^2}
       \left(1-e^{-2\alpha(r)}\right)\right]
\end{multline*}
or
\begin{equation*}
    \frac{1}{r}B^{-1/a}(8\pi)^{1-1/a}=2e^{-2\alpha(r)}\alpha'(r)
      +\frac{1}{r}\left(1-e^{-2\alpha(r)}\right).
\end{equation*}
Rearranging, we obtain
\[
   \frac{2e^{-2\alpha(r)}\alpha'(r)}{B^{-1/a}(8\pi)^{1-1/a}-
      \left(1-e^{-2\alpha(r)}\right)}=\frac{1}{r}
\]
and
\[
   \frac{2\alpha'(r)}{e^{2\alpha(r)}\left[B^{-1/a}(8\pi)^{1-1/a}
     -1\right]+1}=\frac{1}{r}.
\]
The integral formula
\[
    \int\frac{du}{Ae^u+1}=u-\text{ln}\,(Ae^u+1)
\]
now yields
\begin{equation*}
   2\alpha(r)-\text{ln}\left \{e^{2\alpha(r)}\left[B^{-1/a}(8\pi)^{1-1/a}
    -1\right]+1\right \}=\text{ln}\,cr
\end{equation*}
and
\begin{equation*}
   cr\left \{e^{2\alpha(r)}\left[B^{-1/a}(8\pi)^{1-1/a}-1\right]+1
      \right\}=e^{2\alpha(r)},
\end{equation*}
whence
\[
   e^{-2\alpha(r)}=-\left[B^{-1/a}(8\pi)^{1-1/a}-1\right]+\frac{1}{cr}.
\]
Since $\lim_{r \to r_0+}\alpha(r)=+\infty,$ $e^{-2\alpha(r_0)}=0.$  Thus
\[
    c=\frac{1}{r_0}\left[B^{-1/a}(8\pi)^{1-1/a}-1\right]^{-1}.
\]
Substituting $c$ results in
\begin{multline*}
    e^{-2\alpha(r)}=\\-\left[B^{-1/a}(8\pi)^{1-1/a}-1\right]
    +\frac{r_0}{r}\left[B^{-1/a}(8\pi)^{1-1/a}-1\right].
\end{multline*} 
So
\begin{equation}\label{E:ealpha1}
   e^{2\alpha(r)}=\frac{1}{\left[1-B^{-1/a}(8\pi)^{1-1/a}\right]
     \left(1-\frac{r_0}{r}\right)}.
\end{equation}
In the special case of phantom energy, $a=1,$ we have $K(r)=B$, and Eq.~
(\ref{E:ealpha1}) reduces to Eq.~(\ref{E:Zaslavskii}). 

Next, we need to check the flare-out conditions at the throat \cite{MT88}:  
$b(r_0)=r_0$ and $b'(r_0)<1.$  (The requirement of asymptotic flatness, 
$b(r)/r\rightarrow 0$ will be dealt with separately.)  From 
Eq.~(\ref{E:shape1}), 
\begin{equation}\label{E:shape2}
   b(r)=r\left\{1-\left[1-B^{-1/a}(8\pi)^{1-1/a}\right]       
       \left(1-\frac{r_0}{r}\right)\right\}.
\end{equation} 
Evidently, $b(r_0)=r_0.$  The other requirement is 
\begin{equation}\label{E:bprime}
   b'(r_0)=B^{-1/a}(8\pi)^{1-1/a}<1.
\end{equation}

\subsection{Generalized phantom energy}\label{SS:phantom}\noindent
For the generalized phantom energy case, $a>0.$  So raising each side to 
the power $-a$ in Eq.~(\ref{E:bprime}) reverses the sense of the 
inequality.  It follows that
\begin{equation}\label{E:phantominequality}
    B>\frac{1}{(8\pi)^{1-a}},\quad a>0.
\end{equation}
In the special case $a=1,\, K=B>1,$ as required in the phantom-energy 
model.

From Eqs.~(\ref{E:rho}) and (\ref{E:bprime}) 
we can confirm that the WEC is violated at the throat $r=r_0$:
\begin{multline*}
   \rho+p=\rho(1-Br_0^{2(a-1)}\rho^{a-1})\\
    =\rho\left[1-Br_0^{2(a-1)}
   \left(\frac{B^{-1/a}(8\pi)^{1-1/a}}{8\pi r_0^2}\right)^{a-1}\right]\\
   =\rho(1-B^{1/a}(8\pi)^{1/a-1})\\
    <\rho\left(1-\frac{1}{(8\pi)^{1/a-1}}(8\pi)^{1/a-1}\right)=0,
\end{multline*}
since $B>1/(8\pi)^{1-a},\, a>0.$

While the flare-out conditions are met, the spacetime itself is not 
asymptotically flat.  As a result, the wormhole material must be cut off 
at some $r=A$ and joined to the exterior Schwarzschild spacetime
\begin{multline}\label{E:line5}
   ds^2=-\left(1-\frac{2M}{r}\right)dt^2+\frac{dr^2}{1-\frac{2M}{r}}\\
   +r^2(d\theta^2+\text{sin}^2\theta\,d\phi^2).
\end{multline}
Such matching requires continuity of the metric.  As noted in Ref.~\cite{LLO}, 
since the components $g_{\hat{\theta}\hat{\theta}}$ and  
$g_{\hat{\phi}\hat{\phi}}$ are already continuous due to the spherical 
symmetry, one needs to impose continuity only on the remaining components at 
$r=A$:  
 \[
  g_{\hat{t}\hat{t}\text{(int)}}(A)=
      g_{\hat{t}\hat{t}\text{(ext)}}(A) \quad \text{and} \quad 
  g_{\hat{r}\hat{r}\text{(int)}}(A)=
      g_{\hat{r}\hat{r}\text{(ext)}}(A)
\]
for the interior and exterior components, respectively.  These 
requirements, in turn, imply that  
\[
  \gamma_{\text{int}}(A)=\gamma_{\text{ext}}(A)\quad \text{and}\quad 
     b_{\text{int}}(A)=b_{\text{ext}}(A).
\]
In particular, at $r=A,$
\[
    e^{2\alpha(r)}=\frac{1}{1-\frac{b(A)}{A}}=\frac{1}{1-\frac{2M}{A}}.
\]
So we need to determine $M=\frac{1}{2}b(A),$ the total mass of the wormhole for $r\le A$:
\begin{equation*}
   M=\frac{1}{2}A\left\{1-\left[1-B^{-1/a}(8\pi)^{1-1/a}\right] 
     \left(1-\frac{r_0}{A}\right)\right\}.      
\end{equation*}
Since $\gamma(r)=\frac{1}{2}\text{ln}\,\frac{c}{r},$ we have 
\[
   e^{2\gamma(r)}=e^{\text{ln}\,(c/r)}=\frac{c}{r},\quad \text{so that}
       \quad \frac{c}{A}=1-\frac{2M}{A}.
\]
After solving for $c$, we find that
\begin{multline*}
  e^{2\gamma(r)}=\\\frac{1}{r}\left\{A-A\left(1-\left[1-B^{-1/a}(8\pi)^{1-1/a}\right] 
       \left(1-\frac{r_0}{A}\right)\right)\right\}
\end{multline*}
for $r_0\le r\le A.$  At $r=A, \,e^{2\gamma(A)}=1-b(A)/A$, as required.  So for 
$r>A$, the metric becomes 
\begin{multline*}
   ds^2=-\left(1-\frac{b(A)}{r}\right)dt^2+\left(1-\frac{b(A)}{r}\right)^{-1}dr^2\\
   +r^2(d\theta^2+\text{sin}^2\theta\,d\phi^2).
\end{multline*}

\subsection{Generalized Chaplygin gas}\noindent  
For the Chaplygin gas case we have 
\[
   p=-Br^{2(a-1)}\rho^a,\quad  a<0.  
\]
To determine the condition on $B$, it is best to return to Eq.~(\ref{E:ealpha1}) 
and observe that 
\begin{equation}\label{E:inequalitychaplygin1}
   B^{-1/a}(8\pi)^{1-1/a}<1.
\end{equation}
Since $-a>0,$ raising each side to the power $-a$ retains the sense of the 
inequality.  The result is 
\begin{equation}\label{E:inequalitychaplygin2}
    B<\frac{1}{(8\pi)^{1-a}}, \quad a<0.
\end{equation}
This condition is similar to $K<(8\pi r_0^2)^{-(a+1)},\,a>0,$ in the original 
GCG model.  In fact, if $B$ is replaced by $B^{*}r_0^{-2(a-1)}$ in the equation of 
state, then from Eq.~(\ref{E:inequalitychaplygin1}),
\[
     B^{*}<\frac{1}{(8\pi r_0^2)^{1-a}},\quad a<0.
\] 

It also follows from inequality (\ref{E:inequalitychaplygin2}) that the WEC is 
violated at the throat (Sec. \ref{SS:phantom}).

Conversely, if $1/K=B^{-1/a}(8\pi)^{1-1/a}$ in Eq.~(\ref{E:ealpha1}), then 
$B=K^a(8\pi)^{a-1}$, which agrees with Eq.~(\ref{E:ES1}).

\section{The next two solutions}\noindent
As noted earlier, there are only two ways to insert $\gamma(r)$ by hand.  
The second is to assume that $\gamma'\equiv 0$.  If $K$ is constant, 
this yields Lobo's solution \cite{fL05}
\begin{equation}\label{E:Lobo}
   e^{2\alpha(r)}=\frac{1}{1-\left(\frac{r_0}{r}\right)^{1-1/K}}.
\end{equation}
Returning to Eq.~(\ref{E:equation1}), we now have 
\begin{multline}\label{E:equation2}
    \frac{1}{r^2}\left(1-e^{-2\alpha(r)}\right)=Br^{2(a-1)}(8\pi)^{1-a}\\
    \times\left[\frac{2}{r}e^{-2\alpha(r)}\alpha'(r)+\frac{1}{r^2}
       \left(1-e^{-2\alpha(r)}\right)\right]^a.
\end{multline}

Proceding as before, we can put Eq.~(\ref{E:equation2}) into the following form:
\begin{multline}\label{E:equation3}
   \frac{2e^{-2\alpha(r)}\alpha'(r)}
    {1-e^{-2\alpha(r)}}\\
  \times\frac{1}
   {-1+B^{-1/a}(8\pi)^{1-1/a}\left(1-e^{-2\alpha(r)}\right)^{1/a-1}}
       =\frac{1}{r}.
\end{multline}
Integrating the left side requires the formula
\[
   \int\frac{du}{u(-1+Au^B)}=\frac{1}{B}
   \left[\text{ln}(Au^B-1)-\text{ln}\,u^B\right].
\]
The result is 
\[
   \left\{B^{-1/a}(8\pi)^{1-1/a}-\left(1-e^{-2\alpha(r)}\right)^{1-1/a}
    \right\}^{a/(1-a)}=cr.
\]
By letting $r=r_0$, we obtain 
\[
   c=\frac{1}{r_0}\left[B^{-1/a}(8\pi)^{1-1/a}-1\right]^{a/(1-a)}
\]
and, finally,
\begin{equation}\label{E:ealpha2}
   e^{2\alpha(r)}=
   \frac{1}{1-\left[F(r)\right]^{a/(a-1)}},
\end{equation}
where
\begin{multline*}
F(r)=\\B^{-1/a}(8\pi)^{1-1/a}
   \left[1-\left(\frac{r_0}{r}\right)^{(a-1)/a}\right]
   +\left(\frac{r_0}{r}\right)^{(a-1)/a}.
\end{multline*}
In the generalized phantom energy case observe that solution (\ref{E:ealpha2}) 
looks substantially different from solution (\ref{E:Lobo}).  It can be shown, 
however, that 
\begin{equation}\label{E:Lobolimit}
  \lim_{a \to 1}e^{2\alpha(r)}=\frac{1}{1-\left(\frac{r_0}{r}\right)^{1-1/B}};
\end{equation} 
(recall that $B=K$ when $a=1$.)  The reason for this outcome can be seen by 
means of the following heuristic argument: in Eq.~(\ref{E:equation3}), for 
$a$ very close to unity, let $\Omega=\left(1-e^{-2\alpha(r)}\right)^{1/a-1},$ 
so that $\Omega\approx 1$.  Treating $\Omega$ as a constant, 
Eq.~(\ref{E:equation3}) can be integrated to yield
\begin{equation}\label{E:ealpha3}
    e^{2\alpha(r)}=
         \frac{1}
      {1-\left(\frac{r_0}{r}\right)^{1-B^{-1/a}(8\pi)^{1-1/a}\Omega}}.
\end{equation} 
If $a=1$, so that $\Omega=1$, we obtain solution (\ref{E:Lobolimit}).

From 
\begin{multline}\label{E:shape3}
   b(r)=r\left\{B^{-1/a}(8\pi)^{1-1/a}
      \left[1-\left(\frac{r_0}{r}\right)^{(a-1)/a}\right]\right.\\
    \left.+\left(\frac{r_0}{r}\right)^{(a-1)/a}\right\}^{a/(a-1)}
\end{multline}
one can show that 
\[
    b(r_0)=r_0\quad \text{and}\quad b'(r_0)=B^{-1/a}(8\pi)^{1-1/a},
\]
as before [Eq.~(\ref{E:bprime})].  So $b'(r_0)<1$ provided that 
\[
    B>\frac{1}{(8\pi)^{1-a}},\quad a>0.
\]
For the corresponding Chaplygin case,
\[
   B<\frac{1}{(8\pi)^{1-a}},\quad a<0.
\]
As a consequence, the WEC is violated at the throat in both cases, already 
shown in Sec. \ref{S:first}.

For the junction at $r=A$, we determine $b(A)=2M,$ as before.  Then the metric 
becomes
\begin{multline*}
  ds^2=-\left(1-\frac{b(A)}{A}\right)dt^2+e^{2\alpha(r)}dr^2\\
    +r^2(d\theta^2+\text{sin}^2\theta\,d\phi^2)  
\end{multline*}
for $r_0\le r\le A,$ and 
\begin{multline*}
    ds^2=-\left(1-\frac{b(A)}{r}\right)dt^2
      +\left(1-\frac{b(A)}{r}\right)^{-1}dr^2\\
     +r^2(d\theta^2+\text{sin}^2\theta\,d\phi^2)  
\end{multline*}
for $r>A$.

Conversely, suppose $b(r)$ has the form 
\begin{multline*}
   b(r)=r\left\{\frac{1}{K}
      \left[1-\left(\frac{r_0}{r}\right)^{(a-1)/a}\right]\right.\\
    \left.+\left(\frac{r_0}{r}\right)^{(a-1)/a}\right\}^{a/(a-1)},
\end{multline*}
based on Eq.~(\ref{E:shape3}). From Eq.~(\ref{E:shape1}),
\begin{multline*}
    p=-\frac{1}{8\pi r^2}\left\{\frac{1}{K}
      \left[1-\left(\frac{r_0}{r}\right)^{(a-1)/a}\right]\right.\\
    \left.+\left(\frac{r_0}{r}\right)^{(a-1)/a}\right\}^{a/(a-1)}. 
\end{multline*}
Then, using $\rho=b'(r)/(8\pi r^2),$ the expression for 
$-K(r)=p/\rho^a$ can be 
reduced to the form
\[
   K(r)=K^a(8\pi)^{a-1}r^{2a-2},
\]
in agreement with Eq.~(\ref{E:ES1}).


\section{Conclusion}\noindent
This paper introduces a new equation of state with a variable parameter: 
\[
 p=-Br^{2(a-1)}\rho^a, \quad a\neq 0,\quad B>0.
\]
This equation of state generalizes the notion of phantom energy.  The result is 
a combination of the generalized phantom energy and the generalized Chaplygin 
gas models (themselves generalized) to describe a traversable wormhole geometry.  
The traversability conditions can be determined from the WEC violation in 
Sec. \ref{SS:phantom}:
\[
   \rho\left(1-B^{1/a}(8\pi)^{1/a-1}\right)<0
\] 
whenever $B^{1/a}>1/(8\pi)^{1/a-1}.$  This inequality implies the following: if
\[
   B>\frac{1}{(8\pi)^{1-a}},\quad\text{then}\quad a>0,
\] 
and if
\[
   B<\frac{1}{(8\pi)^{1-a}},\quad\text{then}\quad a<0.
\]
In the first case the wormhole is supported by generalized phantom energy and in 
the second by generalized Chaplygin gas.

The solutions obtained are exact, yielding easily stated conclusions, but 
extracted at a price: the redshift function must be inserted ``by hand."  Yet 
apart from the ability to find exact solutions, there does not appear to be 
anything artificial or even special about these redshift functions.  Instead of 
producing purely mathematical solutions, it seems more likely that the 
generalized phantom energy and Chaplygin gas models can, at least in principle, be 
unified by equations of state having a variable parameter.

\end{document}